\documentclass[aps,pra,twocolumn,groupedaddress]{revtex4-2}
\usepackage{graphicx}% Include figure files
\usepackage{amsmath}

\begin{document}

% Use the \preprint command to place your local institutional report
% number in the upper righthand corner of the title page in preprint mode.
% Multiple \preprint commands are allowed.
% Use the 'preprintnumbers' class option to override journal defaults
% to display numbers if necessary
%\preprint{}

%Title of paper
\title{Super-resolution quantum imaging via multiphoton correlation}

% repeat the \author .. \affiliation  etc. as needed
% \email, \thanks, \homepage, \altaffiliation all apply to the current
% author. Explanatory text should go in the []'s, actual e-mail
% address or url should go in the {}'s for \email and \homepage.
% Please use the appropriate macro foreach each type of information

% \affiliation command applies to all authors since the last
% \affiliation command. The \affiliation command should follow the
% other information
% \affiliation can be followed by \email, \homepage, \thanks as well.
% 
% 
% 
% 
% 
\author{Yu Gu}

\author{Yiqi Song}

\author{Meixue Chen}

\author{Yuchen He}

\author{Hui Chen}

\author{Huaibin Zheng}
\affiliation{Electronic Materials Research Laboratory, Key Laboratory of the Ministry of Education \& International Center for Dielectric Research, School of Electronic Science and Engineering, Xi’an Jiaotong University, Xi'an 710049, China}

\author{Yu Zhou}
\affiliation{MOE Key Laboratory for Nonequilibrium Synthesis and Modulation of Condensed Matter, Department of Applied Physics, Xi’an Jiaotong University, Xi’an, 710049, China}

\author{Fuli Li}
\affiliation{MOE Key Laboratory for Nonequilibrium Synthesis and Modulation of Condensed Matter, Department of Applied Physics, Xi’an Jiaotong University, Xi’an, 710049, China}

\author{Zhuo Xu}
\affiliation{Electronic Materials Research Laboratory, Key Laboratory of the Ministry of Education \& International Center for Dielectric Research, School of Electronic Science and Engineering, Xi’an Jiaotong University, Xi'an 710049, China}

\author{Jianbin Liu}
\email[]{liujianbin@xjtu.edu.cn}
\affiliation{Electronic Materials Research Laboratory, Key Laboratory of the Ministry of Education \& International Center for Dielectric Research, School of Electronic Science and Engineering, Xi’an Jiaotong University, Xi'an 710049, China}

%Collaboration name if desired (requires use of superscriptaddress
%option in \documentclass). \noaffiliation is required (may also be
%used with the \author command).
%\collaboration can be followed by \email, \homepage, \thanks as well.
%\collaboration{}
%\noaffiliation

\date{\today}

\begin{abstract}
The resolution of conventional imaging is fundamentally constrained by the diffraction limit. Here we demonstrate that this limit can be overcome using classical thermal light, through the correlation of multiphoton subsystems. By resolving the photon numbers detected in the two arms of a Hanbury Brown–Twiss interferometer, the correlation function exhibits a narrowed peak or dip whose width is tunable via the average photon number. Temporal and spatial super-resolution enhancements up to 11.8-fold and 5.1-fold, respectively, are experimentally achieved. As a proof-of-principle experiment, a double-pinhole that is completely unresolved by standard ghost imaging becomes clearly distinguishable via the correlation of two-photon versus zero-photon. The principle of this super-resolution imaging can be interpreted as a combination of classical super-resolution imaging and quantum super-resolution imaging. This work provides a versatile platform for super-resolution imaging and lithography, and opens a new avenue for exploring the boundary between classical and nonclassical phenomena.
\end{abstract}

% insert suggested keywords - APS authors don't need to do this
%\keywords{}

%\maketitle must follow title, authors, abstract, and keywords
\maketitle

% body of paper here - Use proper section commands
% References should be done using the \cite, \ref, and \label commands
% 
\section{Introduction}
The spatial resolution of an optical imaging system is limited by the wavelength of light and the numerical aperture of the system as $0.61\lambda/NA$, where $\lambda$ and $NA$ are the wavelength and the numerical aperture, respectively \cite{born2021principles}. In order to increase the resolution, shorter wavelengths and larger numerical apertures are required. However, the minimum resolution that can be obtained via this method is limited by the technical difficulties in fabricating short-wavelength light sources and large-numerical-aperture lenses. Different super-resolution imaging techniques \cite{giovannetti2004quantum,hell2015nobel,prakash2022super,aflalo2024optical,yue2025quantum} have been introduced to solve this problem, where the resolution of the imaging system can exceed  $0.61\lambda/NA$. Based on the working principles, super-resolution imaging techniques can be categorized into classical super-resolution imaging \cite{hell2015nobel,prakash2022super,aflalo2024optical} and quantum super-resolution imaging \cite{giovannetti2004quantum,yue2025quantum}. Classical super-resolution techniques are mainly based on single-molecular localization microscopy or point-spread function (PSF) manipulation, which includes photoactivated localization microscopy (PALM) \cite{betzig2006imaging},  stochastic optical reconstruction microscopy (STORM) \cite{rust2006sub}, stimulated emission depletion microscopy (STED) \cite{hell1994breaking}, structured illumination microscopy (SIM) \cite{gustafsson2000surpassing} and so on. Taking STED as an example, the reason why the resolution of STED can surpass $0.61\lambda/NA$ is that stimulated emission is employed to inhibit the fluorescence process in the outer regions of the excitation PSF. The final resolution equals the difference between two PSFs with different widths. Super-resolution can be obtained by suitably choosing the shapes and widths of these two PSFs. A 4.5 times higher resolution was predicted in the original theoretical paper \cite{hell1994breaking} and a 1.5 times higher resolution was experimentally verified later \cite{klar1999subdiffraction}. It was recently reported that 10 times resolution enhancement can be obtained with STED \cite{gopinath2026recent}. Reviews of the classical super-resolution imaging techniques can be found in Refs. \cite{hell2015nobel,prakash2022super,aflalo2024optical}.

Quantum super-resolution imaging is mainly based on that the de Broglie wavelength of $N$ photons is $\lambda/N$, where $\lambda$ is the wavelength of a single photon. The resolution can therefore exceed $0.61\lambda/NA$ by a factor of $N$ \cite{giovannetti2004quantum,yue2025quantum}.  Entangled photons are usually employed to realize quantum super-resolution imaging. Only two times resolution enhancement has been reported in quantum super-resolution imaging with entangled photons \cite{tsang2009quantum,shin2011quantum,unternahrer2018super,toninelli2019resolution,tenne2019super,defienne2021quantum,he2023quantum,zhang2024quantum}, because it is difficult to generate entangled states with more than two photons \cite{hubel2010direct,pan2012multiphoton}. There have been some studies on employing classical light to realize super-resolution quantum imaging or quantum lithography \cite{peer2004quantum,hemmer2006quantum,khoury2006nonlinear,thiel2007quantum,oppel2012superreolving,tsang2016quantum}.   
Hemmer \textit{et al.} showed that it is possible to observe interference pattern with laser light at submultiples of the Rayleigh limit, by employing two laser light beams with specially designed frequencies and a frequency-selective measurement process \cite{hemmer2006quantum}. Oppel \textit{et al.} observed interference patterns with a resolution down to one quarter of the Rayleigh limit using five detectors and independent thermal light sources \cite{oppel2012superreolving}. Khoury \textit{et al.} employed laser light and photon-number projection measurement to observe interference patterns that are a factor of 5 smaller than the Rayleigh limit by superimposing interference patterns with different phase shifts \cite{khoury2006nonlinear}. Special conditions on the detections or the algorithms are required to obtain super-resolution quantum imaging or quantum lithography with classical light \cite{peer2004quantum,hemmer2006quantum,khoury2006nonlinear,thiel2007quantum,oppel2012superreolving,tsang2016quantum}.

In this paper, we propose a new quantum super-resolution imaging method with classical light, which combines the advantages of classical super-resolution imaging (STED) and quantum super-resolution imaging (De Broglie wavelength of multiphoton). This method employ the correlation of multiphoton subsystems of thermal light to retrieve the image of the object. Recently, the study of multiphoton correlations in classical light has attracted considerable attentions, offering the potential to uncover additional correlation properties beyond conventional measurements \cite{cao2024quantum,dawkins2024quantum,you2024isolating,mostafavi2025multiphoton,gu2026observation,chen2026ghost}. Cao \textit{et al.} employed the correlation of multiphoton subsystems of classical light to realize quantum projection ghost imaging with improved visibility and contrast-to-noise ratio  \cite{cao2024quantum}. Mostafavi \textit{et al.} employed natural light to realize quantum imaging with an improved signal-to-noise ratio \cite{mostafavi2025multiphoton}. We have studied the correlation of multiphoton subsystems of thermal light in detail and found that the correlation of multiphoton subsystems of thermal light can be tuned by varying the average number of photons \cite{gu2026observation,chen2026ghost}. The correlation of certain multiphoton subsystems of thermal light is equivalent to a sum of an anti-correlation dip and a correlation peak with different widths. The width of the final observed multiphoton correlation dip can be tuned by varying the average number of photons. Temporal and spatial super-resolution enhancements up to 11.8-fold and 5.1-fold, respectively, are experimentally achieved with our method. A double-pinhole that is completely unresolved by standard ghost imaging becomes clearly distinguishable via the correlation of multiphoton subsystems, which confirms that super-resolution has been achieved.

\section{Theory}

Unlike classical super-resolution imaging, where the intensity is employed to form the image \cite{hell2015nobel,prakash2022super,aflalo2024optical}, quantum super-resolution imaging usually employ the second- or higher-order correlation of light \cite{giovannetti2004quantum,yue2025quantum,peer2004quantum,hemmer2006quantum,khoury2006nonlinear,thiel2007quantum,oppel2012superreolving,tsang2016quantum}. In imaging using the second- or higher-order correlation, the image resolution is determined by the full-width at half maximum (FWHM) of the correlation \cite{shih2021introduction}. For instance, in thermal light ghost imaging with the second-order correlation, the spatial resolution is determined by the FWHM of the normalized second-order coherence function $g^{(2)}(\vec{r}_1,t_1; \vec{r}_2,t_2)$ at equal time \cite{scarcelli2004experimental,martienssen1964coherence}, where $(\vec{r}_1,t_1)$ and $(\vec{r}_2,t_2)$ are the space-time coordinates of photon detection events at detectors 1 and 2, respectively. The FWHM of $g^{(2)}(\vec{r}_1,\vec{r}_2)$ of thermal light equals $\lambda L/D$, where $\lambda$ is the wavelength, $L$ is the distance between the source and detection planes, and $D$ is the diameter of the circular thermal light source \cite{shih2021introduction}. The resolution $\lambda L/D$ is the limit for imaging with the second-order correlation of thermal light, which is equivalent to $0.61\lambda/NA$ in classical imaging. Super-resolution imaging is obtained when the resolution exceeds $\lambda L/D$ in imaging using the second-order correlation, just as it is obtained when the resolution exceeds $0.61\lambda/NA$ in classical imaging. Both limits are set by the diffraction limit of light.

By analogy with thermal light ghost imaging \cite{shih2021introduction}, the quantum imaging with the correlation of multiphoton subsystems of thermal light can be expressed as 
\begin{equation}\label{gi}
Q(x_2)=\int_{S}|O(x_1)|^2g^{(2)}_{mn}(x_1,x_2)dx_1,
\end{equation}
where $O(x_1)$ is the aperture function of the object, $g^{(2)}_{mn}(x_1,x_2)$ is the one-dimension second-order correlation function of detector 1 detecting $m$ photons at $x_1$ and detector 2 detecting $n$ photons at $x_2$, and $S$ is the integral area. All the photons passing through the object are collected by a bucket detector. If the distance between the source and detector 1 planes equals the distance between the source and detector 2 planes, the image can be reconstructed by employing the correlation of multiphoton subsystems \cite{chen2026ghost}. The FWHM of  $g^{(2)}_{mn}(x_1,x_2)$ peak/dip is the resolution of the imaging system, which can be measured in a Hanbury Brown-Twiss (HBT) interferometer with photon-number-resolving detectors \cite{gu2026observation,brown1974intensity}. 

For thermal light in a HBT interferometer, the correlation function $g^{(2)}_{mn}(x_1,x_2)$ can be expressed as \cite{gu2026observation,chen2026ghost}
\begin{eqnarray}\label{gmn}
g^{(2)}_{mn}(x_1,x_2)&=&\frac{(1+R\langle n \rangle )^{m+1}(1+T\langle n \rangle)^{n+1}}{C^{m+n+1}}\\
&& \times \sum_{k=0}^{\min(m,n)}\binom{m}{k}\binom{n}{k}A^{m-k}B^{n-k} \mu^k, \nonumber
\end{eqnarray}
where $A = 1 + (1-\mu)T\langle n \rangle$, $B = 1 + (1-\mu)R\langle n \rangle$, $C = 1 + \langle n \rangle + (1-\mu)RT\langle n \rangle^2$, $\langle n \rangle$ is the average number of photons of the incident thermal light, $R$ and $T$ are the reflectivity and transmissivity of the beam splitter, respectively, and $R+T=1$. $\mu=|g(x_1,t_1;x_2,t_2)|^2$,  where $g(x_1,t_1;x_2,t_2)$ is the first-order normalized correlation function of light field at space-time coordinates $(x_1,t_1)$ and $(x_2,t_2)$. Substituting Eq. (\ref{gmn}) into Eq. (\ref{gi}), an image of the object can be obtained if  $g^{(2)}_{mn}(x_1,x_2)$ reaches its maximum or minimum when $x_1$ equals $x_2$ \cite{chen2026ghost}.

It is complicated to discuss the properties of $g^{(2)}_{mn}(x_1,x_2)$ for general values of $m$ and $n$. Hence, $g^{(2)}_{m0}(x_1,x_2)$ is taken as an example in the following discussions. Substituting $n=0$ into Eq. (\ref{gmn}) and using the expressions for $A$, $B$ and $C$, $g^{(2)}_{m0}(x_1,x_2)$ can be simplified as
\begin{eqnarray}\label{gm0}
&&g^{(2)}_{m0}(x_1,x_2)\\
&=& \frac{(1+R\langle n \rangle)^{m+1}(1+T\langle n \rangle)\bigl[1+(1-\mu)T\langle n \rangle \bigr]^{\,m}}{\bigl[1+\langle n \rangle+(1-\mu)RT\langle n \rangle^2\bigr]^{\,m+1}},\nonumber
\end{eqnarray}
which corresponds to that there are $m$ photons detected in the reflected beam and zero photons detected in the transmitted beam. Equation (\ref{gm0}) can be further simplified when $x_1=x_2$ ($\mu=1$)
\begin{equation}\label{gm0-0}
g^{(2)}_{m0}(0)= \frac{(1+R\langle n \rangle)^{m+1}(1+T\langle n \rangle)}{\bigl[1+\langle n \rangle]^{\,m+1}}.
\end{equation}
For symmetric beam splitter ($R=T=0.5$), $g^{(2)}_{m0}(0)$ attains its minimum value when $\bar{n}$ equals $m$. For positive integer $m$, the minimum value of $g^{(2)}_{m0}(0)$ is less than 1, which indicates that the detection events of $m$ photons and zero photons are anti-correlated \cite{gu2026observation}. $g^{(2)}_{m0}(0)$ tends to 1 in the limit $\bar{n} \to 0$, which indicates that these two detection events are independent in this limit. $g^{(2)}_{m0}(0)$ tends to infinity in the limit $\bar{n} \to \infty$, which indicates that these two detection events are positively correlated.  Based on the analysis above, it can be concluded that as $\bar{n}$ increases from 0, $g^{(2)}_{m0}(0)$ decreases from 1 and reaches its minimum when $\bar{n}$ equals $m$ ($m \geq 1$). As $\bar{n}$ continues to increase from $m$, $g^{(2)}_{m0}(0)$ increases from its minimum and approaches infinity when $\bar{n} \to \infty$. The final $g^{(2)}_{m0}(x_1,x_2)$ can be interpreted as a sum of an anti-correlation dip and a correlation peak. The anti-correlation dip originates from the two-photon bunching of thermal light, and the width of the dip is determined by the properties of thermal light. The correlation peak originates from the multiphoton components, and the width of the peak is determined by the properties of thermal light and the de Broglie wavelength of multiphoton components. If the FWHMs of the dip and peak are different, the FWHM of the final correlation can be much less than the FWHM of the dip or peak. The principle is the same as that of STED \cite{hell1994breaking} except that the shapes and widths of the correlation dip and peak cannot be varied independently in our method.

\begin{figure}[htb]
\centering
\includegraphics[width=85mm]{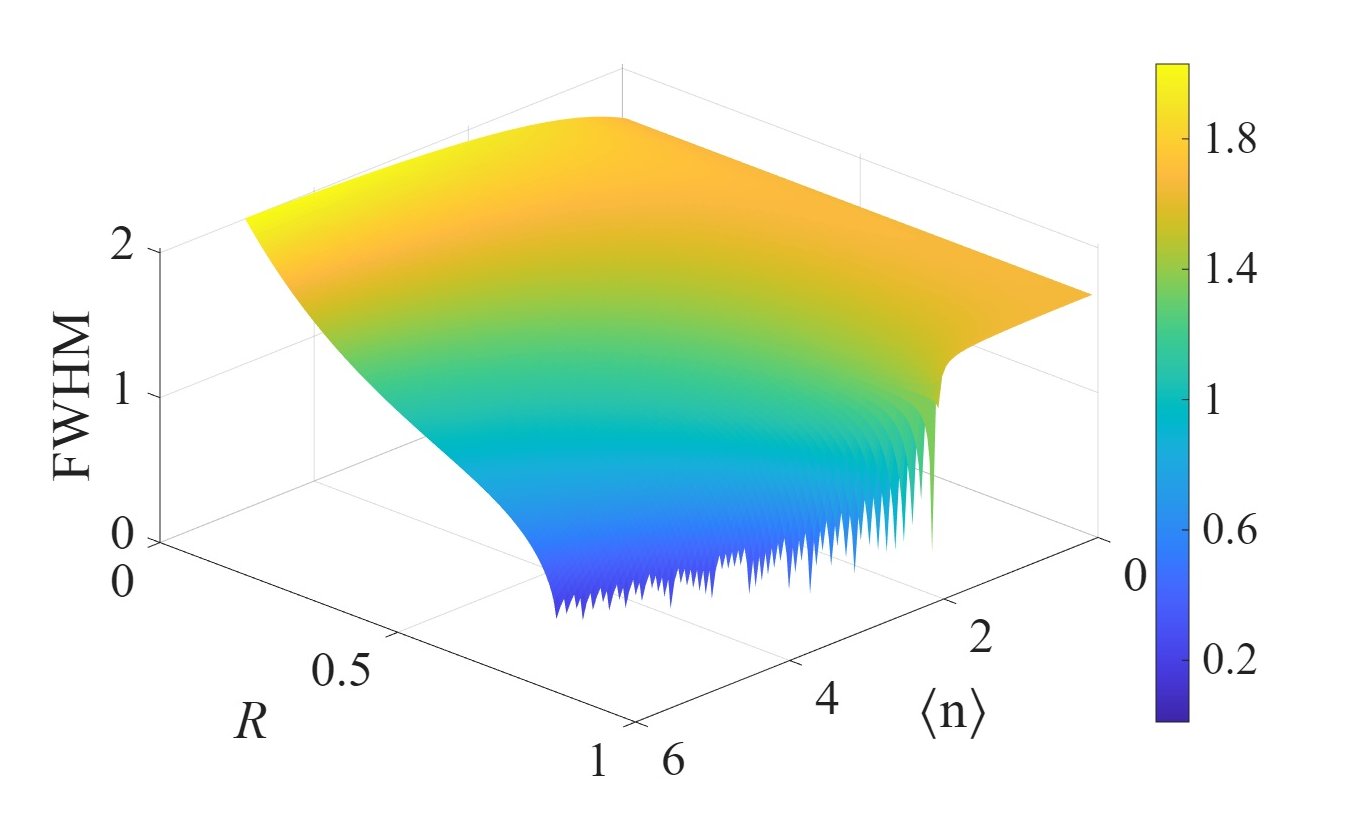}
\caption{Simulated FWHMs of $g^{(2)}_{20}(\tau)$ with Gaussian correlation function. $\tau=t_1-t_2$. $R$ is the reflectivity of the beam splitter. $\langle n \rangle$ is the average number of photons. FWHM of the second-order temporal correlation function of thermal light equals $2\sqrt{2\ln 2} (\approx 2.35) $. This simulation is valid for the temporal correlation and can be generalized to the spatial case.}\label{1-fwhm}
\end{figure}

Figure \ref{1-fwhm} shows the simulated FWHMs of $g^{(2)}_{20}(\tau)$ with the first-order normalized correlation function being Gaussian, $g(\tau)=\exp (-\tau^2/2)$, where $\tau$ is the time difference between the detection events at detectors 1 and 2. For simplicity, the standard deviation of the Gaussian function is set to 1. The FWHM of the second-order temporal correlation function equals $2\sqrt{2\ln 2} (\approx 2.35) $, which is the coherence time thermal light \cite{shih2021introduction}. The FWHM of $g^{(2)}_{20}(\tau)$ is numerically determined by finding the value $\tau_0$ such that $g^{(2)}_{20}(\tau_0)=[1+g^{(2)}_{20}(0)]/2$. Then the FWHM equals $2|\tau_0|$. The FWHM exists only when $g^{(2)}_{20}(0)$ is the minimum of $g^{(2)}_{20}(\tau)$. Hence, there is no FWHM for some of the parameter values in Fig. \ref{1-fwhm}. As shown in Fig. \ref{1-fwhm}, the FWHM of $g^{(2)}_{20}(\tau)$ is always less than  $2\sqrt{2\ln 2}$, which indicates that super-resolution is always realized by observing the correlation of detector 1 detecting two photons and detector 2 detecting zero photons. By properly choosing the values for $R$ and $\langle n \rangle$, the FWHM can be much less than $2\sqrt{2\ln 2}$. The simulations in Fig. \ref{1-fwhm} and discussions above can be generalized to the spatial case.

\begin{figure}[htb]
\centering
\includegraphics[width=85mm]{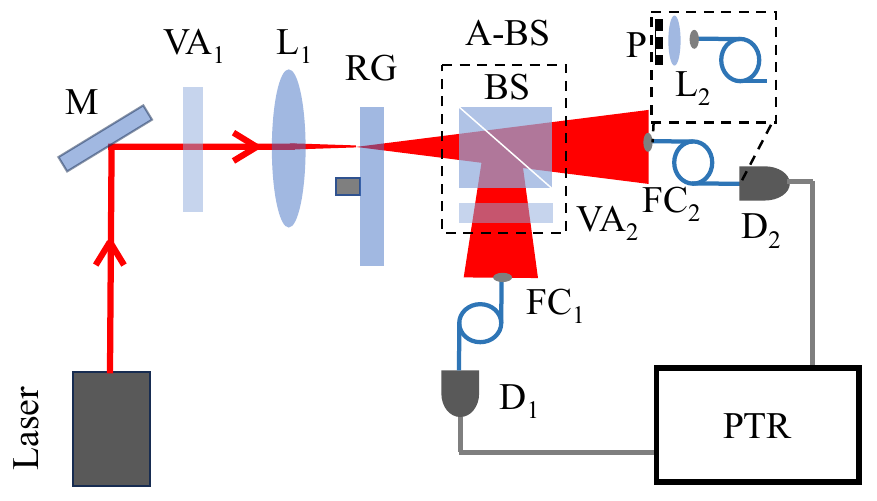}
\caption{Experimental setup to observe super-resolution quantum imaging. Laser: Single-mode continuous-wave laser. M: Mirror. VA$_1$ and VA$_2$: Variable attenuators. L$_1$ and L$_2$: Lenses. A-BS: Asymmetric beam splitter. BS: $1:1$ beam splitter. FC$_1$ and FC$_2$: Fiber couplers. P: double pinholes. D$_1$ and D$_2$: Single-photon detectors. PTR: Photon detection time series recording system. For quantum imaging,  double pinholes and a collecting lens are added before FC$_2$. All the photons passing through the double pinholes are collected by D$_2$.}\label{2-experiment}
\end{figure}

\section{Experiments}

Figure \ref{2-experiment} shows the experimental setup to observe super-resolution quantum imaging. A single-mode continuous-wave laser with a central wavelength of 780 nm and a frequency bandwidth of 20 kHz is employed as the light source. Pseudothermal light is created by focusing the laser light onto a rotating groundglass (RG). The coherence properties of pseudothermal light are controlled by the distance between the lens L$_1$ and RG, the position of laser spot on the RG, and the rotation speed of RG \cite{martienssen1964coherence}. The intensity of the incident light is controlled by a variable attenuator (VA$_1$).  After passing through the RG, the light is split into two beams by an asymmetric beam splitter (A-BS), which is composed of a $1:1$ beam splitter (BS) and a variable attenuator (VA$_2$). For measuring the FWHM of temporal or spatial correlation function, two single-mode fibers with core diameter of 4 $\mu$m are employed to collect the photons. For quantum imaging, a double pinhole and a collecting lens shown in the inset of Fig. \ref{2-experiment} are employed. FC$_1$ and FC$_2$ are two fiber holders. The outputs of these two single-photon detectors (D$_1$ and D$_2$) are sent into a photon detection time series recording system (PTR). The correlation of multiphoton detection events is calculated offline. In our experiments, a HydraHarp 400 module is employed as the PTR, which can be replaced by other modules with similar functions.

\begin{figure}[htb]
\centering
\includegraphics[width=75mm]{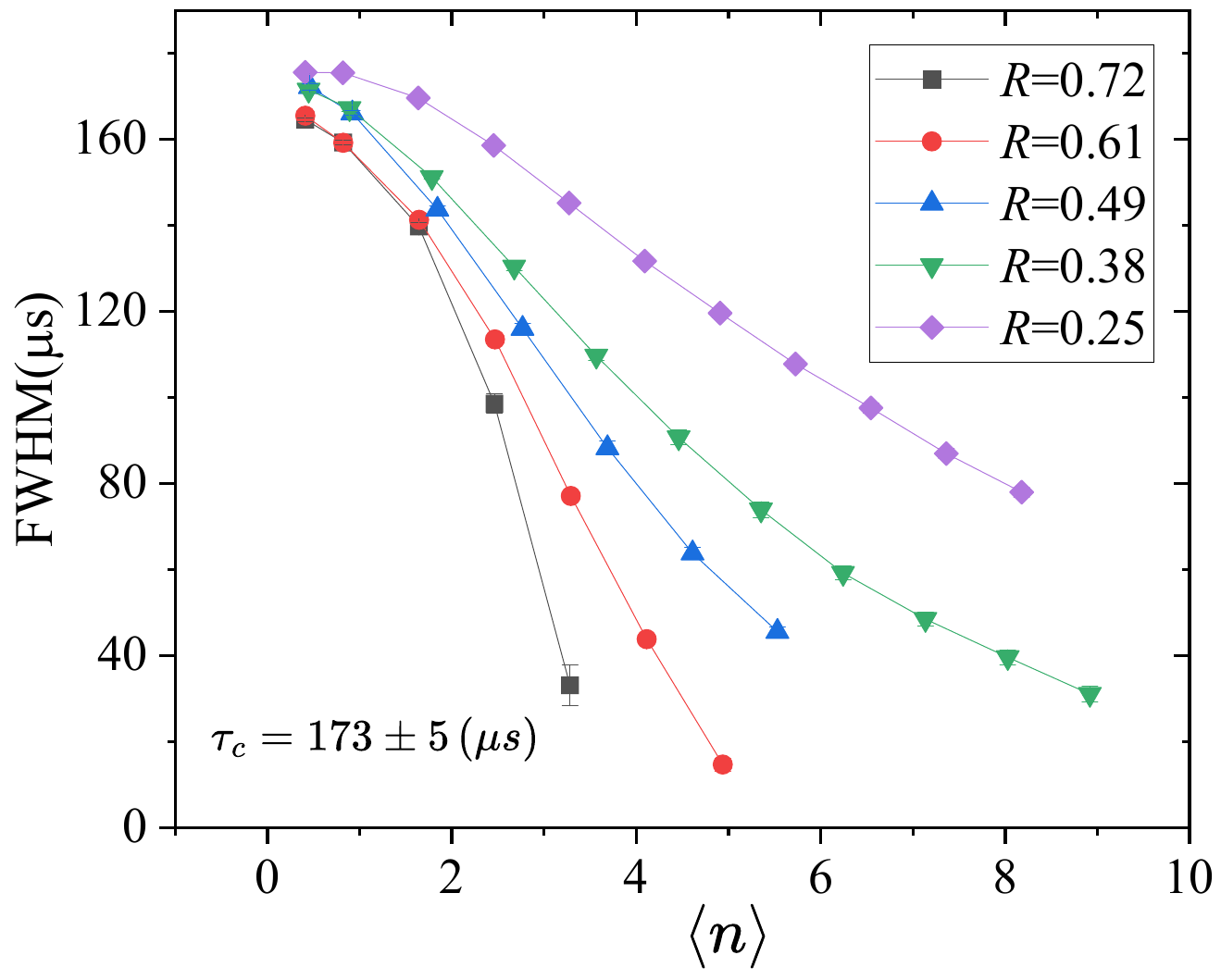}
\caption{Temporal super-resolution via $g^{(2)}_{20}(\tau)$, $\tau=t_1-t_2$. The coherence time of pseudothermal light equals $173 \pm 5$ $\mu$s, which is measured by averaging the FWHMs of the second-order temporal coherence functions with different values of $R$. $R$ is the reflectivity of A-BS. $\langle n \rangle $ is the average number of photons. The shortest FWHM of $g^{(2)}_{20}(\tau)$ equals $14.7 \pm 1.5$ $\mu$s, which represents an 11.8-fold enhancement compared to the coherence time.}\label{3-temporal}
\end{figure}

Figure \ref{3-temporal} shows the measured FWHMs of the temporal correlation function $g^{(2)}_{20}(\tau)$. The coherence time of pseudothermal light equals  $173 \pm 5$ $\mu$s, which is measured by averaging the FWHMs of $g^{(2)}(\tau)$ in different values of $R$. The temporal coherence function is measured when the two detectors are at symmetric positions. For different value of $R$, ten different runs are collected. Each run collects data for 30 s. The total counting rate of the two detectors is at the level of $8 \times 10^5$ c/s. The dark counts of these two detectors are at the level of 60 c/s. The average number of detected photons for a fixed $R$ is calculated by dividing the total counting rate by the width of the time bin. When different time-bin widths are chosen, different average numbers of photons are obtained. As shown in Fig. \ref{3-temporal}, the FWHM of $g^{(2)}_{20}(\tau)$ decreases as $\langle n \rangle $ increases for every $R$. The difference among the $R$ values lies in their decrease rates. Again, the FWHM exists only when $g^{(2)}_{20}(0)$ is the minimum. The minimal FWHM of $g^{(2)}_{20}(\tau)$ is measured to be $14.7 \pm 1.5$ $\mu$s when $\langle n \rangle = 4.94$ and $R=0.61$, which represents an 11.8-fold enhancement compared to the coherence time. By accurately tuning the values of $R$ and $\langle n \rangle$, higher temporal resolution enhancement can be obtained.

\begin{figure}[htb]
\centering
\includegraphics[width=75mm]{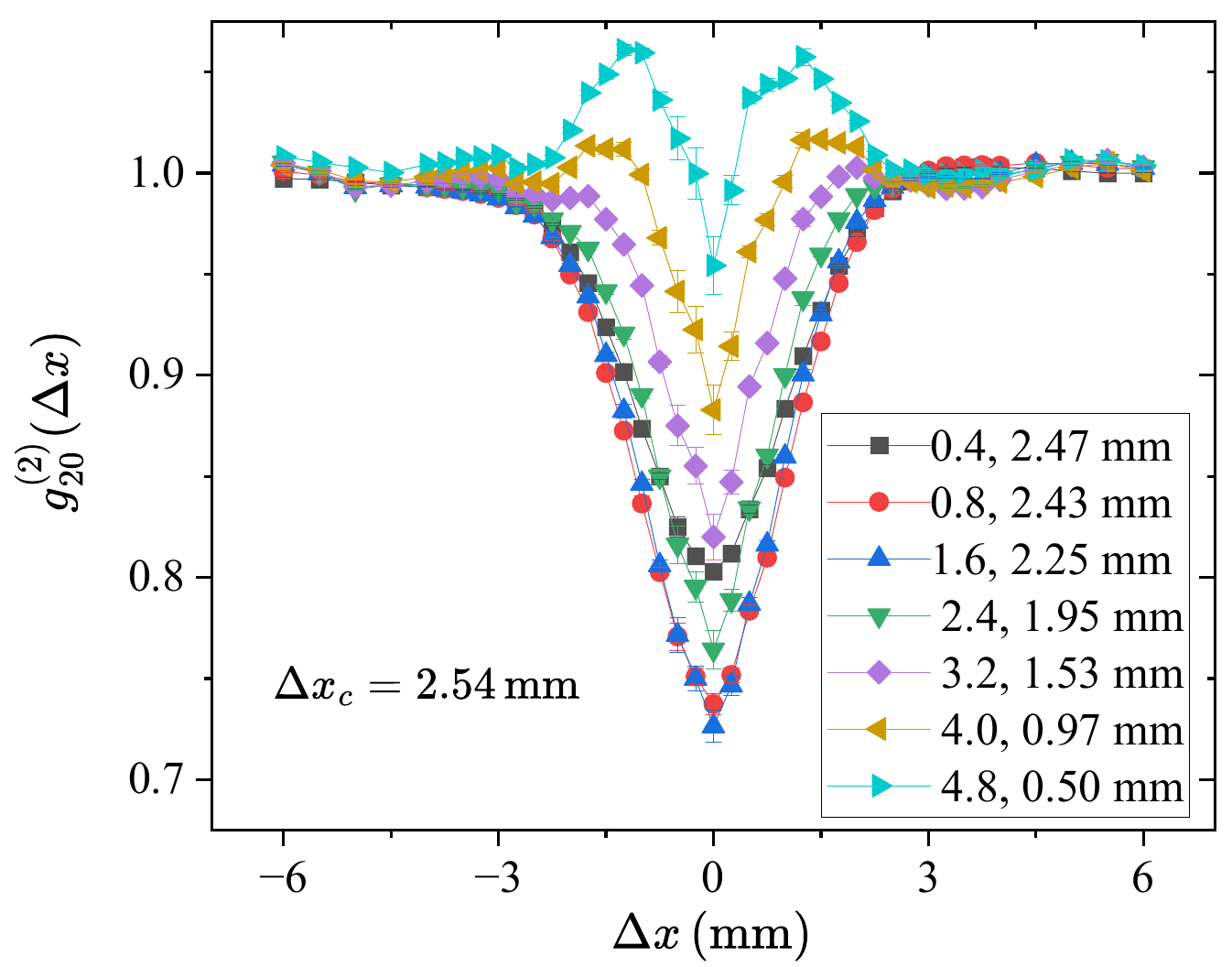}
\caption{Spatial super-resolution via $g^{(2)}_{20}(\Delta x)$, $\Delta x=x_1-x_2$. The transverse coherence length of pseudothermal light in the detection plane equals $2.54$ mm, calculated from the FWHM of the second-order spatial coherence function $g^{(2)}(\Delta x)$. The first number in each legend is the average number $\langle n \rangle$, and the second number is the calculated FWHM of $g^{(2)}_{20}(\Delta x)$  for the corresponding average number of photons.  The minimal FWHM is 0.50 mm, representing a 5.1-fold enhancement compared to the transverse coherence length of pseudothermal light in the detection plane.}\label{4-spatial}
\end{figure}

Figure \ref{4-spatial} shows the measured spatial correlation $g^{(2)}_{20}(\Delta x)$ for different average number of photons, where $\Delta x=x_1-x_2$. The data are obtained by fixing the position of D$_2$ and scanning the position of D$_1$ horizontally. At each scanning position, five independent runs are measured. Each run collects data for 30 s. The counting rates for D$_1$ and D$_2$ equal $(4.9\pm 0.2) \times 10^5$ and $(3.14\pm 0.09) \times 10^5$ c/s, respectively.  The reflectivity of A-BS equals 0.61. The transverse coherence length of pseudothermal light in the detection plane is 2.54 mm, which is calculated via the FWHM of the second-order spatial coherence function $g^{(2)}(\Delta x)$. The FWHM of $g^{(2)}_{20}(\Delta x)$ is less than the transverse coherence length of pseudothermal light, which confirms that super-resolution can be observed via $g^{(2)}_{20}(\Delta x)$ as predicted in Fig. \ref{1-fwhm}. The FWHM equals 0.5 mm for $\langle n \rangle =4.8$, which represents a 5.1-fold enhancement compared to the 2.54 mm coherence length.

\begin{figure}[htb]
\centering
\includegraphics[width=75mm]{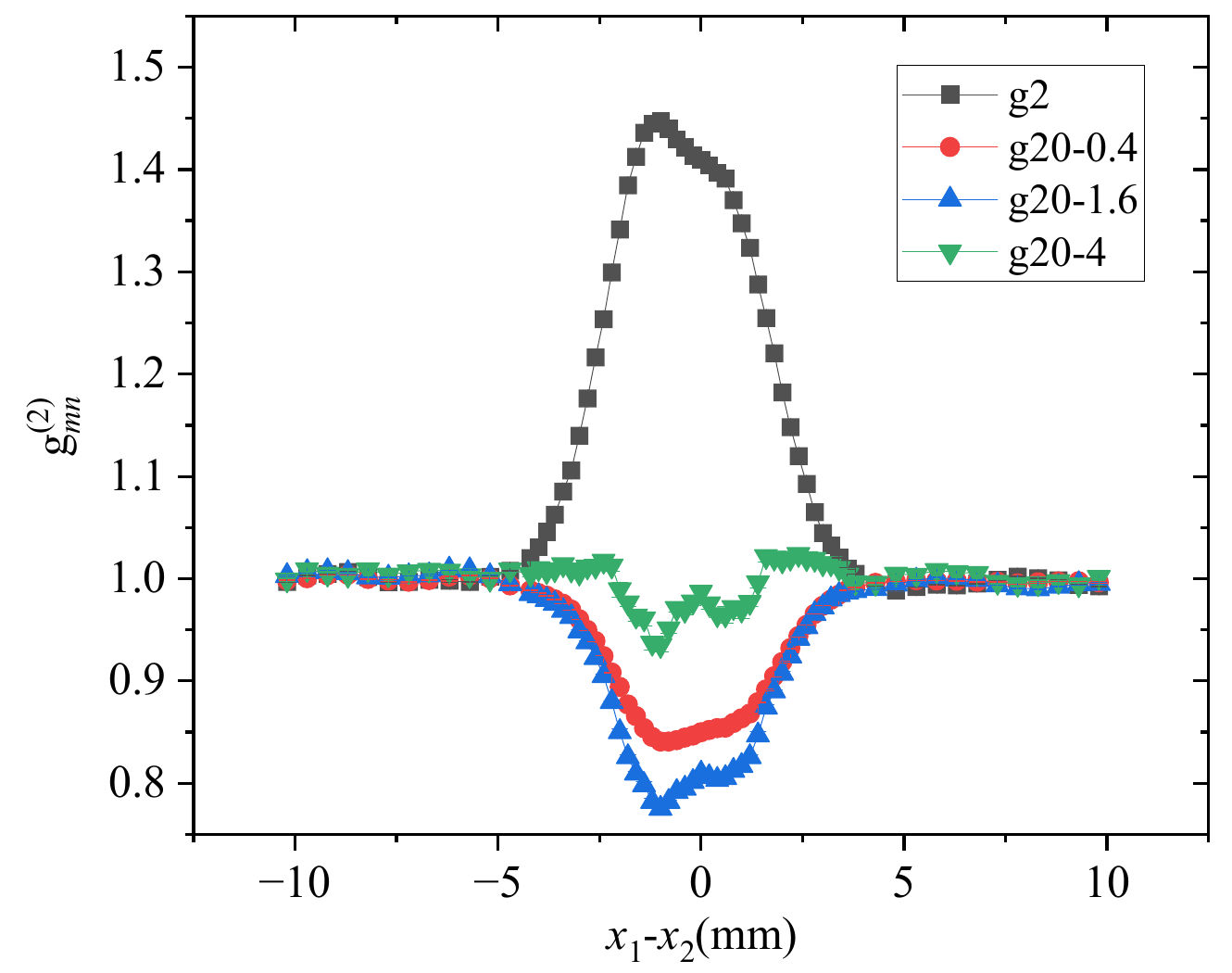}
\caption{One-dimension super-resolution quantum imaging of double pinholes via $g^{(2)}_{20}(0)$. For comparison, the reconstructed image via $g^{(2)}(0)$ is shown by the black squares. $gm0$ is short for $g^{(2)}_{m0}(0)$. The number after $gm0$ is the average number of detected photons. As the average number of photons increases, the structure of double pinholes becomes distinguishable.}\label{5-imaging}
\end{figure}

In order to verify that the observed super-resolution can be employed to realize super-resolution imaging, a double-pinhole is employed as the object in Fig. \ref{2-experiment}. The diameter of each pinhole is 0.1 mm. The distance between the centers of these two pinholes is 2.4 mm. Figure \ref{5-imaging} shows the reconstructed one-dimensional images of the double-pinhole via $g^{(2)}_{20}(0)$. For comparison, the reconstructed image via $g^{(2)}(0)$ is shown by the black squares. No double-pinhole structure is observed via $g^{(2)}(0)$. As the average number of detected photons increases, the structure of double-pinhole becomes obvious, which is consistent with the prediction that the resolution increases as $\langle n \rangle $ increases shown in Fig. \ref{4-spatial}. When $\langle n \rangle =4$, the structure of double-pinhole is clearly observed. The reason why the heights of the correlation dips differ for the two pinholes is that the numbers of photons passing through the two pinholes are different. Even though the quality of the reconstructed image is poor, the results shown in Fig. \ref{5-imaging} can be employed to prove that a structure that cannot be observed with the normal ghost imaging method can be identified with the correlation of multiphoton subsystems. Hence, super-resolution image is realized.

\section{Discussions and conclusions}

Only one-dimension spatial super-resolution imaging is reported. Extending our results to two-dimension spatial imaging \cite{chen2026ghost} and temporal imaging \cite{ryczkowski2016ghost} is straightforward, but beyond the scope of this paper. Khoury \textit{et al.} also employed classical light and photon-number projection measurement to observe interference exceeding the Rayleigh limit \cite{khoury2006nonlinear}. However, their super-resolution interference pattern is observed by superposing patterns with different phase shifts. In our method, only the correlation of multiphoton subsystems is required to obtain super-resolution imaging. Besides, super-resolution temporal imaging can be realized with our method, while the existing studies have concentrated on spatial domain \cite{giovannetti2004quantum,hell2015nobel,prakash2022super,aflalo2024optical,yue2025quantum,peer2004quantum,hemmer2006quantum,khoury2006nonlinear,thiel2007quantum,oppel2012superreolving,tsang2016quantum}. 

Even though the scheme and the algorithm are simple to implement for super-resolution imaging, our method has some limitations. For instance, the measured temporal and spatial super-resolution enhancements are 11.8 and 5.1 times, respectively. However, such enhancement is difficult to achieve in imaging, as shown in Fig. \ref{5-imaging}. The reason behind this phenomenon can be understood by examining the spatial correlations in Fig. \ref{4-spatial}. The FWHM is indeed narrower as the average number of detected photons increases, which indicates that the resolution increases. However, the background becomes larger than 1 as the FWHM becomes narrower. It will cause the measured correlation to change for nearby points. There is a trade-off between super-resolution and image quality.  This limitation can be addressed by specially designing the structure of the correlation function through manipulating thermal light source, or by choosing suitable values of $m$ and $n$ in the correlation $g^{(2)}_{mn}$.

In conclusion, super-resolution imaging is obtained by employing classical thermal light and photon-number projection measurements. By choosing a appropriate correlation between $m$ and $n$ photons in thermal light, both temporal and spatial super-resolution can be achieved. As a proof-of-principle experiment, a double-pinhole is employed as the object for imaging. With the ordinary ghost imaging method, the structure of double-pinhole cannot be distinguished, which means that the spatial resolution of the imaging system is larger than the distance between these two pinholes. By employing the correlation of multiphoton subsystems, the structure of double-pinhole becomes distinguishable as the spatial resolution increases. The scheme and algorithm are simpler compared to the existing super-resolution quantum imaging methods. The principle of our super-resolution imaging can be interpreted as a combination of classical super-resolution (STED) and quantum super-resolution imaging (de Broglie wavelength of multiphoton components). Even though there are some limitations, our method may find applications in super-resolution quantum imaging and quantum lithography.

\begin{acknowledgments}
This work was supported by National Training Program of Innovation and Entrepreneurship for Undergraduates, and the Shaanxi Key Research and Development Project (2024CY2-GJHX-89).
\end{acknowledgments}

\bibliography{antibunching.bib}

\end{document}